# Variance, Skewness & Kurtosis: results from the APM Cluster Redshift Survey and model predictions.

E. Gaztañaga[1,2], R.A.C. Croft[1] & G. B. Dalton[1]

[1] *Department of Physics, Astrophysics, University of Oxford, Nuclear & Astrophysics Laboratory, Keble Road, Oxford OX1 3RH*
[2] *CSIC Centre d'Estudis Avançats de Blanes, Astrophysics, 17300 Blanes, Girona, Spain*

**ABSTRACT**
We estimate the variance $\overline{\xi}_2$, the skewness $\overline{\xi}_3$ and the kurtosis $\overline{\xi}_4$ in the distribution of density fluctuations in a complete sample from the APM Cluster Redshift Survey with 339 clusters and a mean depth $\mathcal{D} \sim 250\,h^{-1}\,\mathrm{Mpc}$.[*] We are able to measure the statistics of fluctuations in spheres of radius $R \simeq 5 - 80\,h^{-1}\,\mathrm{Mpc}$, with reasonable errorbars. The statistics in the cluster distribution follow the hierarchical pattern $\overline{\xi}_J = S_J\,\overline{\xi}_2^{J-1}$ with $S_J$ roughly constant, $S_3 \simeq 2$ and $S_4 \sim 8$. We analyse the distribution of clusters taken from N-body simulations of different dark matter models. The results are compared with an alternative method of simulating clusters which uses the truncated Zel'dovich approximation. We argue that this alternative method is not reliable enough for making quantitative predictions of $\overline{\xi}$. The N-body simulation results follow similar hierarchical relations to the observations, with $S_J$ almost unaffected by redshift distortions from peculiar motions. The standard $\Omega = 1$ Cold Dark Matter (CDM) model is inconsistent with either the second, third or fourth order statistics at all scales. However both a hybrid Mixed Dark Matter model and a low density CDM variant agree with the $\overline{\xi}_J$ observations.

## 1 INTRODUCTION

Rich clusters of galaxies can be used to trace the structure of the Universe on very large scales. Although the identification of clusters is more difficult than that of galaxies, clusters have several important advantages as tracers, both in observations and in simulations. Since clusters represent high densities in the galaxy distribution, it is possible to use them to map out the distant Universe in a systematic and economical way. As clusters are associated with high peaks in the mass distribution, cluster correlations evolve very slowly with time, and so the clustering statistics determined from numerical simulations are relatively insensitive to the epoch at which the simulations are analysed (Croft & Efstathiou 1994a). An additional advantage is that clusters are strongly correlated so that even a small sample can be used to estimate the correlation functions.

The 2-point correlation function for galaxy clusters was first analysed in Abell's (1958) catalogue (e.g. Hauser & Peebles 1973, Bahcall & Soneira 1983, Klypin & Kopylov 1983). Recent determinations based on automated cluster catalogues (Dalton et al. 1992, Nichol et al. 1992) and on flux limited samples of X-ray clusters (Romer et al. 1994) give mutually consistent results. The 2-point correlation of the APM (Automatic Plate Measuring machine) Cluster Redshift Survey has already been used to constrain dark matter models (Dalton et al. 1994b).

Previous analysis of 3-point and 4-point statistics, $\xi_3$ and $\xi_4$, of cluster samples have been based on Abell's catalogue and extensions by Abell, Corwin & Olowin (1989) (e.g. Jing & Zhang 1989, Tóth, Hollósi & Szalay 1989, Jing & Valdarnini 1991, Plionis & Valdarnini 1995, Cappi & Maurogordato 1995). These samples seem to reproduce the hierarchical relation $\xi_J \sim \xi_2^{J-1}$ found in the galaxy distribution (e.g., Groth & Peebles 1977; Fry & Peebles 1978; Sharp et al. 1984; Szapudi, Szalay & Boschan 1992; Meiksin, Szapudi & Szalay 1992; Bouchet et al. 1993; Gaztañaga 1992, 1994). These observations could provide important insights into the underlying matter distribution, and the formation of clusters. However, there is evidence that the clustering measured from Abell's catalogue is affected by inhomogeneities in the selection of clusters (e.g. Efstathiou et al. 1992) and it is not clear how this could affect the above mentioned estimates of $\xi_3$ and $\xi_4$.

Here we estimate the variance $\overline{\xi}_2$, the skewness $\overline{\xi}_3$ and the kurtosis $\overline{\xi}_4$ in the distribution of density fluctuations in a sample from the APM Cluster Redshift Survey. The selection of clusters in the APM Cluster Catalogue is based on a computer algorithm applied to digitised photographic plates and so provides a more uniform sample of clusters than Abell's. We compare the results with previous analysis and with different cluster simulations.

[*] Throughout the paper we use $H_0 = 100h\,\mathrm{km/s/Mpc}$.

In next section, we review the APM cluster data and N-body models. We present the method of estimation and the results in Section 3. Sections 4 and 5 are devoted to the discussion and conclusions. In an appendix we consider the properties of simulations generated using the truncated Zel'dovich approximation.

## 2  CLUSTER SAMPLES

### 2.1  The APM Cluster Survey

We use sample B from Dalton et al. (1994b) which contains 364 clusters with mean redshift 0.086 and space-density $3.4 \times 10^{-5}$ $h^3 \mathrm{Mpc}^{-3}$. The clusters were selected from the APM Galaxy Survey (Maddox et al. 1990b,c) using an automatic selection algorithm described in detail by Dalton et al. (in preparation). We have cut the sample to redshifts larger than 10000km/s to avoid uncertainties in the selection function at low redshift, and imposed an upper limit of 40000km/s. There are only 25 clusters outside this range, so that the total number in the selected region is $N = 339$. As the redshifts are obtained from only a few galaxies per cluster, the individual cluster redshifts are uncertain to within 700km/s (Dalton et al. 1994a).

### 2.2  Simulations

The cluster simulations are those generated by Croft & Efstathiou (1994a,b), using a particle–particle–particle-mesh N-body code (Efstathiou & Eastwood 1981, Efstathiou et al. 1985), and are the same as those used by Dalton et al. (1994b). We consider three different models. First, the standard CDM model which is scale-invariant and has $\Omega_0 = 1$ (hereafter $\Gamma \equiv \Omega h = 0.5$, CDM). Second, a scale-invariant, low-density CDM universe with $\Omega_0 = 0.2$, $h = 1$ and a cosmological constant $\Lambda$ such that $\Lambda/3H_0^2 = 1 - \Omega_0$ ($\Gamma = 0.2$, CDM). The power spectrum for these two models is taken from Efstathiou, Bond & White (1992). Third, a scale-invariant model containing both a massive neutrino component (7-eV neutrinos) and CDM, such that $\Omega_{total} = 1.0$, $\Omega_{CDM} = 0.6$, $\Omega_\nu = 0.3$, with $h = 0.5$ (Mixed Dark Matter, MDM) from Klypin et al. (1993). The models have been normalised to be consistent with the amplitude of fluctuations in the microwave background (Wright et al. 1994), so that the value of the linear theory rms mass fluctuations in 8 $h^{-1}$ Mpc spheres, $\sigma_8 = 1.0$ for the two CDM models and 0.67 for MDM. 10 realisations of each model were run, using different random phases, in boxes of side length 300 $h^{-1} \mathrm{Mpc}$, with $10^6$ particles. We have also run 4 realisations of the $\Gamma = 0.2$, CDM with the slightly higher normalisation $\sigma_8 = 1.3$. Clusters of particles are located using a percolation algorithm, the cluster mass is defined within a metric radius and the richness limit of the cluster sample is chosen to match the observed cluster space density. The metric radius is cho-sen to be 0.5 $h^{-1} \mathrm{Mpc}$ and the mean intercluster separation 30 $h^{-1} \mathrm{Mpc}$ (Croft & Efstathiou 1994a).

We have also tested an alternative scheme for making theoretical predictions about cluster clustering. This scheme, advocated by Plionis et al. (1995) makes use of the Zel'dovich (1970) approximation to dynamically evolve the density field before clusters are selected. In order to assess the validity of results derived using this method we have examined clustering in the Zel'dovich approximation and carried out a comparison with N-body clusters. These tests are detailed in Appendix A.

## 3  CLUSTER CORRELATIONS

### 3.1  Method of estimation

We apply a variant of the method used by Efstathiou et al. (1990), which accounts for selection effects automatically by estimating the mean density and its fluctuations in shells of mean constant redshift. We divide space in the redshift catalogue into a series of concentric shells, of width $2R$, centred on the observer, which are further subdivided into $M_s$ spherical cells of radius $R$. For each shell $s$ we compute the mean number of particles $\overline{N}_s(R)$ in a cell and the moments $m_J(R)$:

$$m_J = \sum_{c=1}^{M_s} (N_c - \overline{N}_s)^J, \tag{1}$$

where $N_c$ is the number of galaxies in the $c^{\mathrm{th}}$ cell. The corresponding connected moments $k_J = \overline{N}_s^J \overline{\xi}_J$, corrected for Poisson shot-noise (see section 3.4), are given by:

$$\begin{aligned} k_2 &= m_2 - \overline{N}_s \\ k_3 &= m_3 - 3m_2 + 2\overline{N}_s \\ k_4 &= m_4 - 3m_2^2 - 6m_3 + 11m_2 - 6\overline{N}_s. \end{aligned} \tag{2}$$

(see e.g. Gaztañaga 1994). Thus, $\overline{\xi}_2 = k_2/\overline{N}_s^2$ is the variance (also called $\sigma^2$), $\overline{\xi}_3 = k_3/\overline{N}_s^3$ is the skewness and $\overline{\xi}_4 = k_4/\overline{N}_s^4$ the kurtosis.

To estimate the mean value of $\langle \overline{\xi}_2 \rangle$ for the whole sample, we introduce the standard Gaussian likelihood function, so that the value of $k_2$ in each shell is weighted by the inverse of its variance, Var($k_2$)$_s$:

$$\mathrm{Var}(k_2)_s = \frac{k_4 - 2k_3 + k_2 - k_2^2}{M_s}, \tag{3}$$

where we have made the assumption that the cells are independent. Thus the weighting compensates for the effects of fluctuations from both shot-noise (important for distant, diluted shells) or from having a small number of cells (important in the nearby shells). We have tried two prescriptions to estimate this variance. The first one uses the higher order moments $m_J$ in the shell to estimate $k_J$ in equation (3). In the second prescription we assume a Gaussian distribution to estimate $k_J$ from $m_2$ [as in Efstathiou et al. (1990)]. Both give similar results.

For higher order moments we use the same weights for each shell as the ones used to estimate $\overline{\xi}_2$ so that, in general:

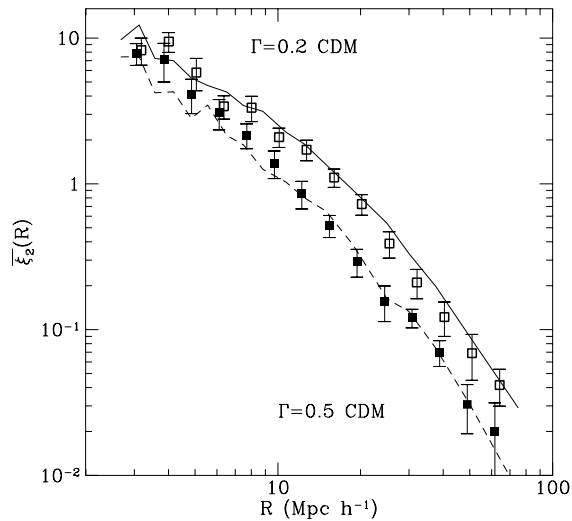

**Figure 1.** The variance $\overline{\xi}_2(R)$ estimated from counts-in-cells in a simulated mock catalogue of clusters (symbols with error-bars), compared with the values from clusters selected from the same realization of the simulation in a box (lines). The dashed line and filled squares correspond to the standard $\Omega = 1$ CDM model whereas solid lines and open squares correspond to low density CDM.

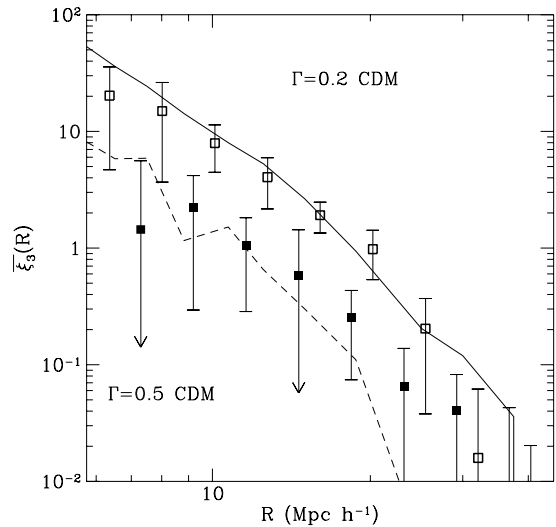

**Figure 2.** The skewness $\overline{\xi}_3(R)$ estimated from counts-in-cells in a simulated mock catalogue of clusters compared with the full simulation, as in Figure 1.

$$\langle \overline{\xi}_J \rangle = \mathcal{K} \sum_{s=1}^{M_s} \frac{\overline{N}_s^J}{\mathrm{Var}(k_2)_s} \overline{\xi}_J, \quad (4)$$

$$\mathcal{K} \equiv \left[ \sum_{s=1}^{M_s} \frac{\overline{N}_s^J}{\mathrm{Var}(k_2)_s} \right]^{-1}$$

where $\overline{\xi}_J = \overline{N}_s^J k_J$ with $k_J$ evaluated for the $s$ shell.

One of the most valuable merits of this approach is that it also provides an estimate of the variance on this mean, $\mathrm{Var}(\overline{\xi}_J)$, as we have all the values in each shell and their relative weights:

$$\mathrm{Var}(\overline{\xi}_J) = \frac{1}{M-1} \langle \overline{\xi}_J^2 \rangle - \langle \overline{\xi}_J \rangle^2, \quad (5)$$

where $M$ is the total number of shells, and the averages use the same weightings as equation (4).

This is repeated for different values of $R$, the end product being the mean $\overline{\xi}_J(R)$ and its variance. The cells in each shell must not overlap too much; this is important because otherwise fluctuations in $M_s$ are not taken into account in the weighting and nearby shells tend to dominate the statistics.

## 3.2 Check of the method using simulated clusters.

We have used the simulations described above to check our method. We first estimate the "true" values of $\overline{\xi}_J(R)$ using counts in cells in redshift-space in the simulation box, in the standard way (see e.g. Baugh, Gaztañaga & Efstathiou 1995). Next we transform the simulations to mimic the geometry of the observational data, applying the APM survey mask and the selection function obtained by Dalton et al. (1994b). The above counts in cells method is used to estimate $\langle \overline{\xi}_J \rangle$ in the mock survey and compare them with results from direct estimation.

In Figure 1 we show the comparison for mock $\Gamma = 0.5$ and $\Gamma = 0.2$ CDM catalogues against the results from the original simulated box. In both cases we have included redshift distortions. Figure 1 shows that there is good agreement, although it should be noted that a perfect match is not expected at large scales because the mock catalogue only samples a fraction of the total volume in the box. These results are for just one simulation (not the average of 10 realizations), showing that it is possible to recover the intrinsic clustering from the catalogues using the above method.

Figure 2 shows the recovered skewness, $\overline{\xi}_3$, from the $\Gamma = 0.5$ and $\Gamma = 0.2$ CDM catalogues compared with the results from the full cubical box. As expected the errors are much larger here. The errors in the $\Gamma = 0.5$ model are even larger and it is difficult to obtain significant results from just one catalogue given that the clustering amplitudes are smaller.

## 3.3 Results for APM clusters and dark matter models

Using the above method, we are able to measure correlations with a very large cell radius, up to $R \simeq 80\, h^{-1}\,\mathrm{Mpc}$. The smallest scales are limited by the shot-noise in the cell occupancy to $R \simeq 5\, h^{-1}\,\mathrm{Mpc}$. The error bars come from equation (5), which provides a consistent estimate of the variance of $\overline{\xi}_J$ within the sample.

To make an accurate estimate of $\overline{\xi}_J$ in the simulations we use all clusters in each cubical box of side $300\, h^{-1}\,\mathrm{Mpc}$. There are $\sim 1000$ clusters in each simulation. In this way the uncertainties are typically smaller than the symbols we use to compare with the data. As mentioned above we have also made mock catalogues of the simulations, with the APM mask and selection function, and find a good agreement, within the errors, between the results from the mock catalogue (using the shell method) and the direct results from the cube (see § 3.2 above).

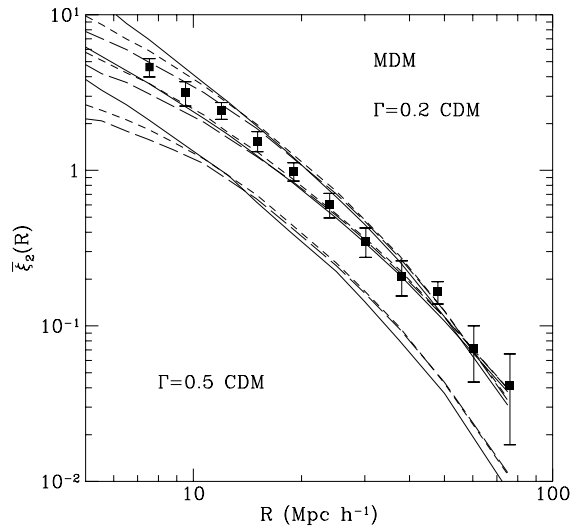

**Figure 3.** The variance $\overline{\xi}_2(R)$ estimated from counts-in-cells in the APM clusters (symbols with with error-bars) compared to the values in different simulations. The continuous, short-dashed and long-dashed lines correspond repectively to the estimation in real space, redshift space and redshift space with Gaussian velocity errors. The lines show the standard $\Omega = 1$ CDM model (lower lines), the low density CDM variant (middle lines) and the MDM model (upper lines).

In Figures 3-5 we show the values of $\overline{\xi}_2(R)$, $\overline{\xi}_3(R)$ and $\overline{\xi}_4(R)$, obtained from the APM clusters (symbols with error bars). The estimates for $J = 4$ become very uncertain, especially at large scales. The "real space" results for $\overline{\xi}_J$ in each model are shown as continuous lines in Figures 3-5. We estimate $\overline{\xi}_J$ for each model after applying redshift distortions using the peculiar velocities for the clusters in the simulations, short-dashed lines in figures. Finally, as the observations are affected by errors in the redshift, we also show, as long-dashed lines, the estimates of $\overline{\xi}_J$ in redshift space after adding errors with a Gaussian distribution of dispersion 700 km/s to the simulated cluster redshifts.

Redshift distortions produce slightly larger clustering amplitudes at large scales, because of coherent infall, and smaller amplitudes at small scales, because of random velocities. The crossing of the real space and the redshift space curves indicates the scale at which both effects are comparable. We expect Gaussian errors of 700 km/s to reduce the redshift space results at small scales, $\sim 700/H_0$, in agreement with the figures.

In Figures 6 and 7 we show the hierarchical ratios $S_J \equiv \overline{\xi}_J / \overline{\xi}_2^{J-1}$ with errors generated from the errors in $\overline{\xi}_J$. The mean values for the APM cluster are around $S_3 \simeq 2$ and $S_4 \sim 8$. At the largest scales both the data and the models have large errors, nevertheless we show the results at these scales to illustrate that even when $R \simeq 50\ h^{-1}$ Mpc there is still a significant detection of $S_3$.

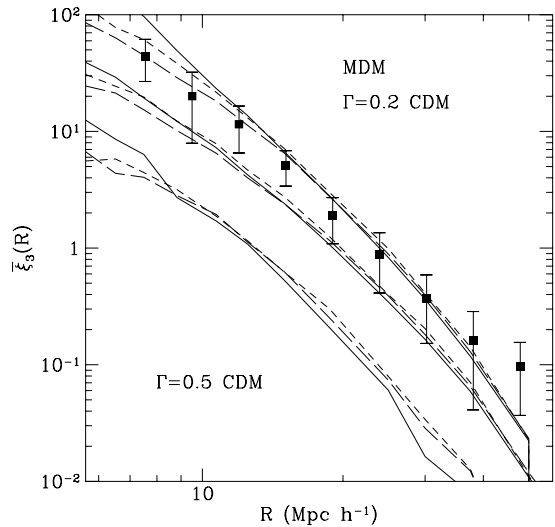

**Figure 4.** The skewness $\overline{\xi}_3(R)$ (symbols with error-bars) estimated from counts-in-cells in the APM clusters compared to the values in different simulated models, as in Figure 3.

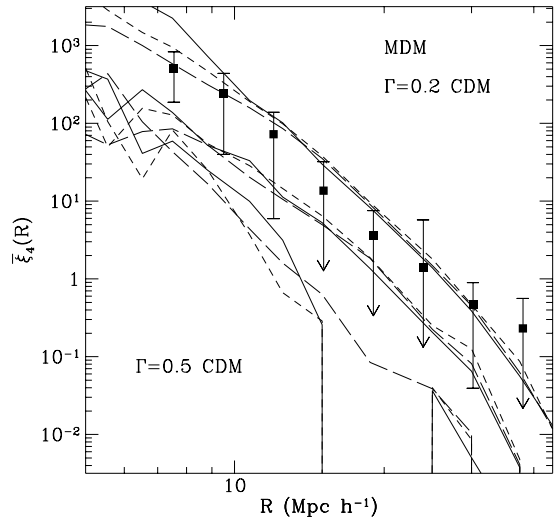

**Figure 5.** The kurtosis $\overline{\xi}_4(R)$ estimated from counts-in-cells in the APM clusters (symbols with error-bars) compared with the values in different simulated models, as in Figure 3.

Note that the simulations show that redshift-space distortions produce only a small effect in $S_J$ as the distortions in $\overline{\xi}_J$ tend to compensate for the distortions in $\overline{\xi}_2^{J-1}$ (Fry & Gaztañaga 1994).

From Figure 3 it is apparent that $\overline{\xi}_2$ in the $\Gamma = 0.2$ model has too little power on scales $R \lesssim 20\ h^{-1}$ Mpc. This is also evident in the skewness $\overline{\xi}_3$, shown in Figure 4. Although we believe this effect is real, it is not very significant if we use $2 - \sigma$ errorbars.

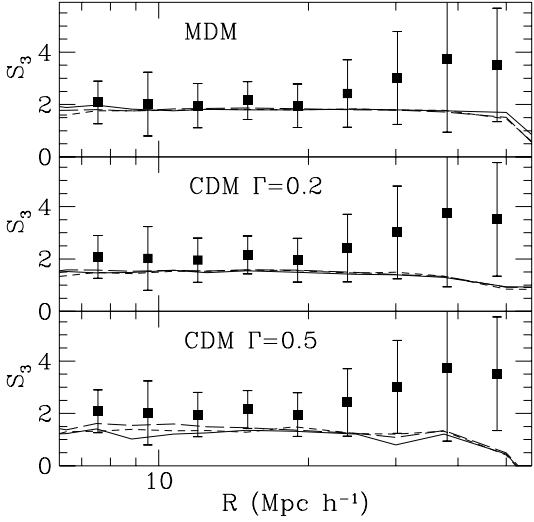

**Figure 6.** The hierarchical skewness $S_3 = \overline{\xi}_3 / \overline{\xi}_2^2$ for the APM clusters (symbols with error-bars) compared with values from different cosmological models. The lower panel shows the low density CDM model, while the upper panel shows the MDM model. The continuous, short-dashed and long-dashed lines correspond repectively to the model estimation in real space, redshift space and redshift space with Gaussian velocity errors.

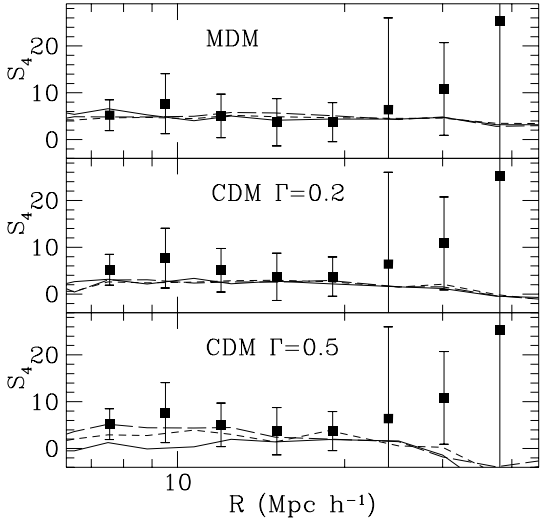

**Figure 7.** The hierarchical kurtosis $S_4 = \overline{\xi}_4 / \overline{\xi}_2^3$ for the APM clusters (symbols with error-bars), compared with the values from different models, as in Figure 6.

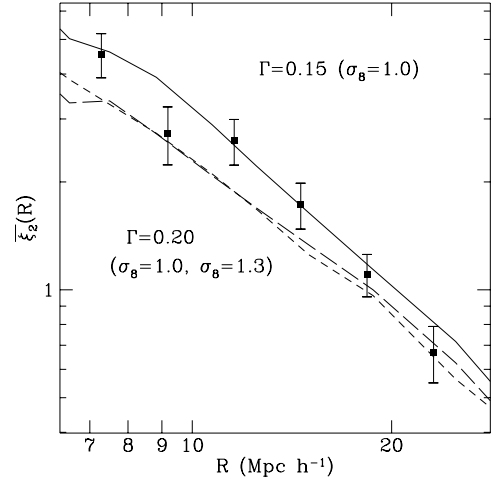

**Figure 8.** The variance $\overline{\xi}_2(R)$ in the $\Gamma = 0.20$ model normalized to $\sigma_8 = 1.0$ (short-dashed line) and $\sigma_8 = 1.3$ (long-dashed line) compared to $\overline{\xi}_2(R)$ in the $\Gamma = 0.15$ model.

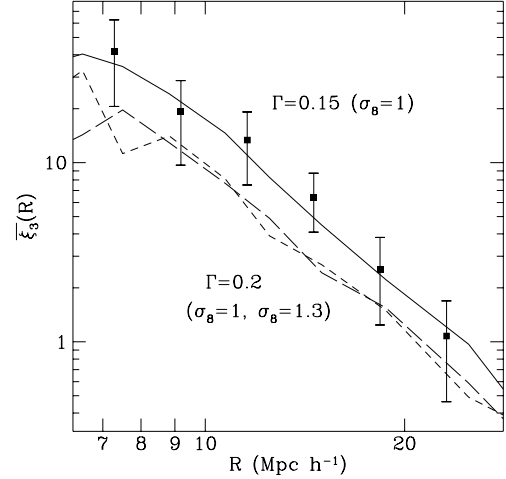

**Figure 9.** The skewness $\overline{\xi}_3(R)$ in the $\Gamma = 0.20$ model normalized to $\sigma_8 = 1.0$ (short-dashed line) and to $\sigma_8 = 1.3$ (long-dashed line) compared to $\overline{\xi}_2(R)$ in the $\Gamma = 0.15$ model (normalized so that $\sigma_8 = 1.0$).

While the amplitude of $\overline{\xi}_2$ is not very sensitive to the normalization of the model, $\sigma_8$, it does depend strongly on the value of $\Gamma$. To show this we have run one realization of a CDM $\Gamma = 0.15$ model (with $\Omega = 0.2$ and $h = 0.75$). This is shown in Figures 8 and 9, where we compare the results of the $\Gamma = 0.15$ and $\Gamma = 0.20$ models, with two normalizations. Because we only have one realization of the $\Gamma = 0.15$ model, these estimates correspond to just one set of random phases (identical for each model). With just one realization it is difficult to distinguish between these three models on scales larger than $20 \, h^{-1}$ Mpc. The results at intermediate scales have sufficiently small errors to constrain the shape parameter $\Gamma$ to have a value of $\sim 0.15$. This is in good agreement with results from the APM galaxy angular correlation function (Maddox et al. 1990a, Efstathiou, Maddox & Sutherland 1992).

### 3.4  Discreteness

In both the observed cluster catalogues and the cluster simulations the distribution of clusters is represented by a discrete number of points. The main effect of this discreteness is to introduce additional correlations (sometimes called 'shot-noise') which are important at scales $R \lesssim d_n$, where $d_n$ is the mean interparticle separation. Typically, if we have a correlation length $R_0$, which is defined as $\bar{\xi}_2(R_0) = 1$, with $R_0 > d_n$, as in the case of galaxies, then the intrinsic clustering dominates over the shot-noise. For clusters, $R_0 \lesssim d_n$, so the shot-noise dominates the measured clustering.

This problem becomes evident when estimating $\bar{\xi}_J$ from moments of counts in cells because one has to decide explicitly what (if any) discreteness correction should be applied. But the problem is equally important when estimating the 2-point correlation function, $\xi_2$, directly. In the latter case one counts distances between pairs in the sample and compares them with similar counts in a random (Poisson) distribution of points. Thus, the Poisson shot-noise correction has been widely used for clusters even when this is not always explicitly stated.

One could choose not to correct the measured correlations for this discreteness as long as the same density of points is used in the comparison with models and other catalogues. This is also the case if one regards the cluster distribution as intrinsically discrete and richness dependent, i.e. characterised by its mean density and therefore by $d_n$.

On the other hand one could try to correct for discreteness and estimate the intrinsic correlations of an underlying density field. This estimate would be independent of the interparticle separation and would be more useful in comparisons with models or other catalogues.[†] The cluster distribution can then be imagined as a continuous field which is traced by our particular point distribution. This continuous field is not the matter distribution, but a biased transformation of it. As in the case of the galaxy distribution one can further assume that the location of the point drawn from the underlying fluctuations is the result of a (Poisson) random process (with probability proportional to local density). In this sense, the Poisson shot-noise model can be applied to correct for the discreteness.

In our case, the Poisson correction is larger than unity at scales $R \lesssim 19\,h^{-1}$ Mpc, as the mean cluster density, $\overline{N}$, in spheres with $R \simeq 19\,h^{-1}$ Mpc have $\overline{N} = 1$. This can be seen in Figure 10, where the uncorrected values of $\bar{\xi}_2$ for the $\Gamma = 0.5$ model (filled squares) and the $\Gamma = 0.2$ model (filled triangles) are compared with the values of $\bar{\xi}_2$ after the discreteness correction, with open squares (triangles) for the $\Gamma = 0.5$ ($\Gamma = 0.2$) CDM model.

In the same Figure we have also plotted the Poisson shot-noise correction $\bar{\xi}_2 = \overline{N}^{-1} \propto R^{-3}$ as a dashed line. Note that, for the $\Gamma = 0.5$ model, the shot-noise contribution is larger than the intrinsic clustering contribution at *all* scales. Although we know that the intrinsic clustering properties are different in these models, we can see in Figure 10 how the uncorrected values of $\bar{\xi}_2$ approach the Poisson correction on small scales, where discreteness is important. Given that we have not used the Poisson model in deciding how to select clusters in the simulations this result indicates that Poisson shot-noise model is indeed a good approximation to the discreteness effects.

[†] One would then be able to disentangle, for example, the effects of incompleteness (such as random sampling) and intrinsic richness dependence in a catalogue of clusters.

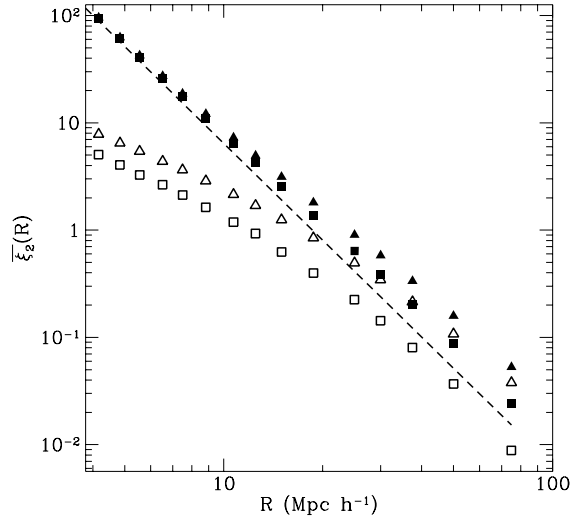

**Figure 10.** The uncorrected values of $\bar{\xi}_2$ for the $\Gamma = 0.5$ model (filled circles) and the $\Gamma = 0.2$ model (filled triangles) compared with the values of $\bar{\xi}_2$ after the discreteness (Poisson shot-noise) correction, with open circles (triangles) for the $\Gamma = 0.2$ ($\Gamma = 0.5$) CDM model. The Poisson shot-noise correction $\bar{\xi}_2 = \overline{N}^{-1} \propto R^{-3}$ is shown as a dashed line.

## 4  DISCUSSION AND CONCLUSIONS

We find two common features in the cluster simulations of different dark matter models. First, redshift distortions are not very important. This is in spite of the fact that different models have quite different velocity fields (e.g. Croft & Efstathiou 1994b). Second, the values of $S_3$ and $S_4$ are quite similar in all models, i.e. in Figures 6 and 7. This is in spite of the fact that different models have quite different values of the variance, skewness or kurtosis (Figures 3-5).

Both of these features can be understood in terms of biasing. Clusters are high peaks in the galaxy (mass) distribution and they are therefore biased tracers of the mass (Kaiser 1984). On large scales one expects coherent streaming motions to enhance redshift space fluctuations (Kaiser 1987), but the relative effect is suppressed (by a factor $b$, the ratio of the mass value to the cluster value of $\bar{\xi}_2$) for a biased distribution. On the other hand the amplitudes of $S_J$ do not seem to be much affected by redshift distortions for either the mass distribution, according to perturbation theory (Bouchet et al. 1994), or for galaxies, according to observations (Fry & Gaztañaga 1994). This might explain why for $\bar{\xi}_J = S_J\,\bar{\xi}_2^{J-1}$, with $J > 2$, redshift distortions are also not very important.

Why should $S_J$ be similar in all models? Following Fry

& Gaztañaga (1993), a local bias relating galaxy or matter fluctuations $\delta$ to cluster fluctuations $\delta_c$:

$$\delta_c(\boldsymbol{x}) = f[\delta(\boldsymbol{x})] = \sum_{k=0}^{\infty} \frac{b_k}{k!} \delta^k(\boldsymbol{x}), \qquad (6)$$

where $b_1 \equiv b$, produces amplitudes for the cluster distribution $S_3$ and $S_4$ of:

$$\begin{aligned} S_3 &= b^{-1}(S_{3,m} + 3c_2) \\ S_4 &= b^{-1}(S_{4,m} + 12c_2 S_{3,m} + 4c_3 + 12c_2^2) \end{aligned} \qquad (7)$$

where $c_2 = b_2/b$ and $c_3 = b_3/b$ are the respective quadratic and cubic contributions from the biasing transformation in equation (6), and $S_{J,m}$ are the hierarchical amplitudes in the underlying matter distribution. Here we have used the fact that both the galaxy distribution in the APM and the matter distribution in the simulations follow the hierarchical relation $\bar{\xi}_J \simeq S_J \bar{\xi}_2^J$ (see e.g. Gaztañaga 1994, Baugh, Gaztañaga & Efstathiou 1995). From the above expression we see that the underlying amplitudes $S_{J,m}$ are typically suppressed by $1/b$ when selecting high peaks. That the values of $S_3$ and $S_4$ we find are quite similar in all models, even when the matter amplitudes $S_{3,m}$ and $S_{4,m}$ are significantly different (see Juszkiewicz, Bouchet & Colombi 1993, Gaztañaga & Baugh 1995) indicates that the biasing contribution in equation (7) is significant. For example, the typical variation between models of $S_{3,m}$ for the mass distribution at large scales is $\Delta S_{3,m} \simeq 2$. Everything being equal, the corresponding variation in the cluster distribution, from equation (7), is $\Delta S_3 \lesssim 0.6$ for the typical value $b \simeq 3$. Given the uncertainties in the measured values of $S_3$ it is not surprising that this small variation is undetected.

Recently, Plionis & Valdarnini (1995) and Cappi & Maurogordato (1995) have estimated $S_3$ and $S_4$ from angular and redshift positions of Abell and ACO clusters. The hierarchial ratios shown in Figures 6 and 7, $S_3 \simeq 2$ and $S_4 \sim 8$ are similar to their redshift-space results, but are slightly smaller than the ones found by Cappi & Maurogordato in the angular distribution, $S_3 \simeq 3$ and $S_4 \sim 15$. It is not clear to us how significant this discrepancy is, as Cappi & Maurogordato (1995) use bootstrap errors, which are unreliable (see Baugh, Gaztañaga & Efstathiou 1995). On the other hand uncertainties from the deprojection of angular into 3-D statistics are not taken into account by Cappi & Maurogordato. This can introduce large errors given the uncertainties in the selection function of clusters.

Plionis et al. (1995) have made predictions based on use of the Zel'dovich approximation to evolve cosmological density fields before selecting clusters. They rule out the $\Gamma = 0.2$, CDM model at a high significance level on the basis that it produces a cluster distribution with too high a variance and skewness on scales $r > 20~h^{-1}$ Mpc. In Appendix A we have carried out tests on this method of simulating the cluster distribution and we find it to be unreliable. Problems arise from the inability of the Zel'dovich approximation to predict accurately the location of the compact objects which we are using to trace the density field. In the case of $\Gamma = 0.2$ CDM we have also run Zel'dovich realisations using the same parameters (particle number, grid spacing, box length and cluster richness) as Plionis et al. and have used the same method to evaluate the moments (from the cluster point distribution smoothed onto a grid with a Gaussian kernel, radius $R$). We find that the variance and skewness are largely overestimated in comparison with the N-body results. In our test, the comparison is in real space and the discrepancy is $\sim 50\%$ for $R = 20$ and $R = 30$ $h^{-1}$ Mpc. This suggests that although Plionis et al. use results from a different observational catalogue of clusters (the combined Abell-ACO catalogue) to make their quantitative comparison, their conclusions are different and in our view less trustworthy than those which would be reached using full N-body simulations.

In summary, we have shown that in the case of cluster samples discreteness has an important effect on clustering statistics at all scales. The Poisson shot-noise model has been widely used implicitly when estimating the 2-point correlation function for clusters. Here we have argued that the Poisson shot-noise correction should indeed be applied when estimating correlations $\bar{\xi}_J$ in the cluster distribution. We reproduce previous conclusions by Dalton et al. (1994b) based on the 2-point correlation function, mainly that the standard $\Omega = 1$ CDM model is inconsistent with the second order statistics at all scales. As a new result we find that this model is inconsistent with either the third or fourth order statistics at all scales. A Mixed Dark Matter model gives good agreement, as does a low density CDM variant, both of these models being consistent with the $\bar{\xi}_J$ observations for $J = 2 - 4$ at the $2\sigma$ level. We have also shown that it is possible to measure $S_3$ in the cluster distribution at scales as large as $R \simeq 50~h^{-1}$ Mpc. The hierarchical correlations found in the cluster distribution are consequence of similar relations in the underlying matter field. In a forthcoming paper, we use the relations between the cluster values of $S_J$ and the correponding values for the matter distribution [e.g. equation (7)] to investigate the biasing process in more detail (in preparation).

## Acknowledgements

We would like to thank George Efstathiou for helpful comments on a previous version of this manuscript. EG acknowledges support of a Fellowship from the Commission of European Communities and from CSIC (Consejo Superior de Investigaciones Científicas) in Spain. RACC acknowledges receipt of a PPARC studentship.

## REFERENCES

Abell, G.O. 1958, ApJ, Supp. 3, 211

Abell, G.O., Corwin, H.G., Olowin, R.P. 1989, ApJ, Supp. 70, 1 (ACO)

Bahcall, N.A., Soneira, R.M., 1983, ApJ, 270, 20

Bardeen, J. M., Bond, J. R., Kaiser, N., & Szalay, A. S. 1986, ApJ, 304, 15

Baugh, C.M., Efstathiou, G., 1993, MNRAS, 265, 145

Baugh, C.M., Gaztañaga, E. Efstathiou, G., 1995, MNRAS, *in press*

Bernardeau, F., 1994a ApJ, 433, 1

Bernardeau, F., 1994b A&A, 291, 697

Bernardeau, F., Kofman 1994 ApJ, *in press*

Bower, R.G., Coles, P., Frenk, C.S., White, S.D.M., 1993, ApJ, 405, 403

Bouchet, F.R., Strauss, M.A., Davis, M., Fisher, K.B., Yahil, A., Huchra, J.P., 1993, ApJ, 417, 36

Bouchet, F.R., Colombi, S., Hivon, E., Juszkiewicz, R., 1994, A&A, submitted

Cappi A., Maurogordato, S. 1995, ApJ, 438, 507

Coles, P., Melott, A.L., Shandarin, S.F., MNRAS, 260, 75

Colombi, S., Bouchet, F.R., Schaeffer, R., 1993, Astron. Astroph., 281, 301

Croft, R.A.C., Efstathiou, G., 1994a, MNRAS, 267, 390

Croft, R.A.C., Efstathiou, G., 1994b, MNRAS, 268, L23

Dalton, G.B., Efstathiou, G., Maddox, S.J., Sutherland, W.J., 1994a, MNRAS, 269, 151

Dalton, G.B., Croft, R.A.C., Efstathiou, G., Sutherland, W.J., Maddox, S.J., Davis, M., 1994b, MNRAS, 271, L47

Dalton, G.B., Efstathiou, G., Maddox, S.J., Sutherland, W.J., 1995, *in preparation.*

Efstathiou, G., Bond, J.R., White, S.D.M, 1992, MNRAS, 258, 1p

Efstathiou, G., Eastwood, J.W., 1981, MNRAS, 194, 503

Efstathiou, G., Davis, M., Frenk, C.S., White, S.D.M., 1985, ApJ Supp.57, 241

Efstathiou, G., Kaiser, N., Saunders, W., Lawrence, A., Rowan-Robinson, M., Ellis, R.S., Frenk, C.S., 1990, MNRAS, 247, 10p

Efstathiou, G., Sutherland, W.J., Maddox, S.J., 1990, Nature 348, 705

Efstathiou, G., Dalton, G.B., Sutherland, W.J., Maddox, S.J., 1992, MNRAS, 257, 125

Frieman, J.A., Gaztañaga, E. 1994, ApJ, 425, 392

Fry, J.N., Peebles, P.J.E. 1978, ApJ, 221, 19

Fry, J.N. 1984, ApJ, 279, 499

Fry, J.N., Gaztañaga, E., 1993, ApJ, 413, 447

Fry, J.N., Gaztañaga, E., 1994, ApJ, 425, 1

Gaztañaga, E., 1992, ApJ, , 398, L17

Gaztañaga, E., 1994, MNRAS, , 268, 913

Gaztañaga, E., Frieman, J.A., 1994, ApJ, 437, L13

Gaztañaga, E., Baugh, C.M., 1995, MNRAS, *in press*

Goroff, M.H., Grinstein, B., Rey, S.J., Wise, M.B., 1986, ApJ, 311, 6

Groth, E.J., Peebles, P.J.E., 1977, ApJ, 217, 385

Hauser, M.G., Peebles, P.J.E., 1973, ApJ, 185, 757

Hockney, R.W., Eastwood, J.W., 1981, Computer Simulations using Particles, McGraw-Hill, New York.

Jing, Y.P., Zhang 1989, ApJ, 342, 639

Jing, Y.P., Valdarnini 1991, A&A, 250, 1

Kaiser, N., 1984, ApJ, 284, L9

Kaiser, N., 1987, MNRAS, 227, 1

Klypin, A.A., Kopylov, A.I., 1983, Sov. Astron. Lett.9, 41

Klypin, A.A., Holtzmann, J., Primack, J., Regös, E., 1993 ApJ, 416, 1

Juszkiewicz, R., Bouchet, F.R., Colombi, S. 1993 ApJ, 412, L9

Maddox, S.J., Efstathiou, G., Sutherland, W.J., Loveday, J., 1990a, MNRAS, 242, 43p

Maddox, S.J., Efstathiou, G., Sutherland, W.J., Loveday, J., 1990b, MNRAS, 243, 692

Maddox, S.J., Efstathiou, G., Sutherland, W.J., Loveday, J., 1990c, MNRAS, 246, 433

Mann, R.G., Heavens, A.F., Peacock, J.A., 1993, MNRAS, 263, 798

Meiksin, A., Szapudi, I. & Szalay, A.S. 1992, ApJ, 394, 87

Nichol, R.C., Collins, C.A., Guzzo, L., Lumsden, S.L., 1992, MNRAS, 255, 21p

Peebles, P.J.E., 1980, The Large Scale Structure of the Universe: Princeton University Press

Plionis, M., Valdarnini, R., 1995, MNRAS, *in press*

Plionis, M., Borgani, S., Moscardini, L., Coles, P., 1995, ApJ, *in press*

Romer, A.K., Collins, C.A., MacGillivray, H., Cruddace, R.G., Ebeling, H., Boringer, H., 1994, Nature, 372, 75

Sathyaprakash, B.S., Sahni, V., Munshi, D., Pogosyan, D., Melott, A.L., 1994, MNRAS, *submitted*

Sharp, N.A., Bonometto, S.A., Lucchin, F. 1984, ApJ, 130, 79

Szapudi, I., Szalay, A.S. & Boschan, P. 1992, ApJ, 390, 350

Tóth, G., Hollósi, J., Szalay, A.S. 1989, ApJ, 344, 75

Weinberg, D.H., Cole, S., 1992, MNRAS, 259, 652

Wright, E.L., Smoot, G.F., Kogut, A., Hinshaw, G., Tenorio, L., Lineweaver, C., Bennett, C.L., Lubin, P.M., 1994, ApJ, 420, 1

Zel'dovich, Ya., B., 1970, A&A, 5, 84

# APPENDIX A: CLUSTERS IN THE ZEL'DOVICH APPROXIMATION

In this appendix we are concerned with testing some aspects of the reliability of theoretical predictions of cluster clustering. In particular we compare our N-body simulations with simulations that use the Zel'dovich (1970) approximation to evolve the density field. Such simulations have been used as a computationally inexpensive way of making predictions about clustering and higher order moments of clusters for a wide range of models (Plionis et al. 1995 and references therein).

## A1 Alternative methods of simulating clusters.

The two-point correlation function of simulated rich clusters evolves weakly with time (Croft & Efstathiou (1994a)). This is because much of the clustering measured is statistical, in the sense that the clusters were born clustered, as would happen if they formed at the locations of high peaks in the initial density field (Kaiser 1984). Less time consuming methods than the use of full N-body simulations might therefore be adequate to model the statistical clustering properties of clusters. Analytical approaches based on high peak theory (eg. Mann, Heavens & Peacock 1993) predict the same trends as N-body simulations – the cluster correlation amplitude increases weakly with cluster richness, for example. To make accurate predictions which can be compared quantitatively with observational samples, though, it is certainly better to identify clusters with peaks in the evolved density field.

This is the view that has been taken by Plionis et al. (1995) in their study of cluster correlations. The dynamical scheme which they use is the Truncated Zel'dovich Approximation (TZA), which has been tested by several authors (e.g. Coles, Melott & Shandarin 1993, Sathyaprakash et al. 1994) and found to give a surprisingly good indication of the behaviour of gravitating matter even into the non-linear regime. It should be noted that agreement found with N-body simulations was only good on scales much larger than the typical radius of galaxy clusters. We also note that it fails to reproduce in detail the amplitudes of the higher order correlations $S_J$ of the mass at all scales (e.g. Juszkiewicz et al. 1993). When the original Zel'dovich (1970) approximation is applied to a particle density field all particles move along their initial trajectories with constant velocity. When particles trajectories cross one another, structure which would have formed under the action of gravity is erased. For this reason, in the improved version, the truncated ZA, the initial power spectrum is truncated (by smoothing with a gaussian filter) to remove the power at small wavelengths which is responsible for the bulk of orbit crossing. Using the TZA to evolve the density field rather than a full N-body code is computationally much cheaper, meaning that large numbers of models can be run and tested. This is potentially of benefit, as for example Plionis et al. (1995) have run simulations of 6 different universes, with which it would be interesting to compare our APM cluster statistics.

It is therefore important to test the reliability of this sort of scheme, particularly as we have reached the stage where cluster samples are large enough that we can measure higher order statistics with relatively small errors. We would like to know whether clusters do indeed form at sites that can be predicted by weakly non-linear evolution of the density field. If this does occur, then it could give support to the idea that statistical clustering (or "high-peak biasing") is mainly responsible for the distribution of clusters. On the other hand, if we are not able to find a reliable correspondence between TZA and N-body clusters the matter distribution may still be similar for both methods when smoothed on large scales, but in the TZA we will have been unable to predict correctly the locations of clusters that we are using to trace that matter distribution.

## A2 Comparison of Truncated Zel'dovich Approximation and N-body clusters.

We test which of these is in fact the case by evolving the matter distribtion for the same initial random phases using the TZA and the $P^3M$ N-body code. In this example case we will use the $\Gamma = 0.2$ CDM models normalised so that $\sigma_8 = 1.3$. The input power spectrum of the TZA run was truncated by use of a Gaussian filter of radius $r_g = 6.3\ h^{-1}$ Mpc which Plionis et al. (1995) find is optimal for this particular model. The final particle distribution for a 15 $h^{-1}$ Mpc thick region is shown in Figure A1b(ii), and can be compared directly with the N-body particle distribution (Figure A1a(ii)). It can be seen that there are broad similarities between the two density fields. There are however fewer obvious clumps in the TZA density field. We are expecting this as small scale power has been removed from the power spectrum and also no mechanism exists within the dynamical scheme to form structure into virialised objects. We must therefore decide where such objects would have formed. Plionis et al. assume that this would occur at peaks in the TZA density field and we will follow them in defining clusters to be at the positions of peaks. However, examination of the TZA particle distribution shows that there are many smooth-looking regions which we would like to break up in order to reproduce the N-body results.

To find clusters we grid the final particle density fields using a TSC scheme (Hockney & Eastwood 1981) onto a $256^3$ mesh and then apply Gaussian smoothing. The positions of the 1000 highest peaks in the smoothed field are taken as the locations of the 1000 richest clusters. This technique does have a number of disadvantages, such as the fact that resolution is lost below the scale of the grid, and also due to the smoothing procedure. There is an additional anti-correlation effect, arising from the fact that peaks cannot lie in adjacent grid-cells. This is discussed further below when we calculate clustering statistics. To check the sensitivity to differences in cluster-finding techniques of the N-body results, we apply the above procedure first to the N-body outputs. We try filters of radius $r_f = 0.5, 1.0$ and $2.0\ h^{-1}$Mpc. As the interparticle separation in cluster cores is small, we should not have to smooth much in order to remove any discreteness effects (in fact using the unsmoothed grid does give good results). In figure A1a(ii) we compare the positions of these N-body clusters (for $r_f = 1.0\ h^{-1}$Mpc) with those identified using percolation. It can be seen that there is an extremely good correspondence between the two cluster catalogues, as we would expect. The panel A1a(ii) shows the distribution of particles taken from the 100 $h^{-1}$Mpc square region indicated in the left hand picture.

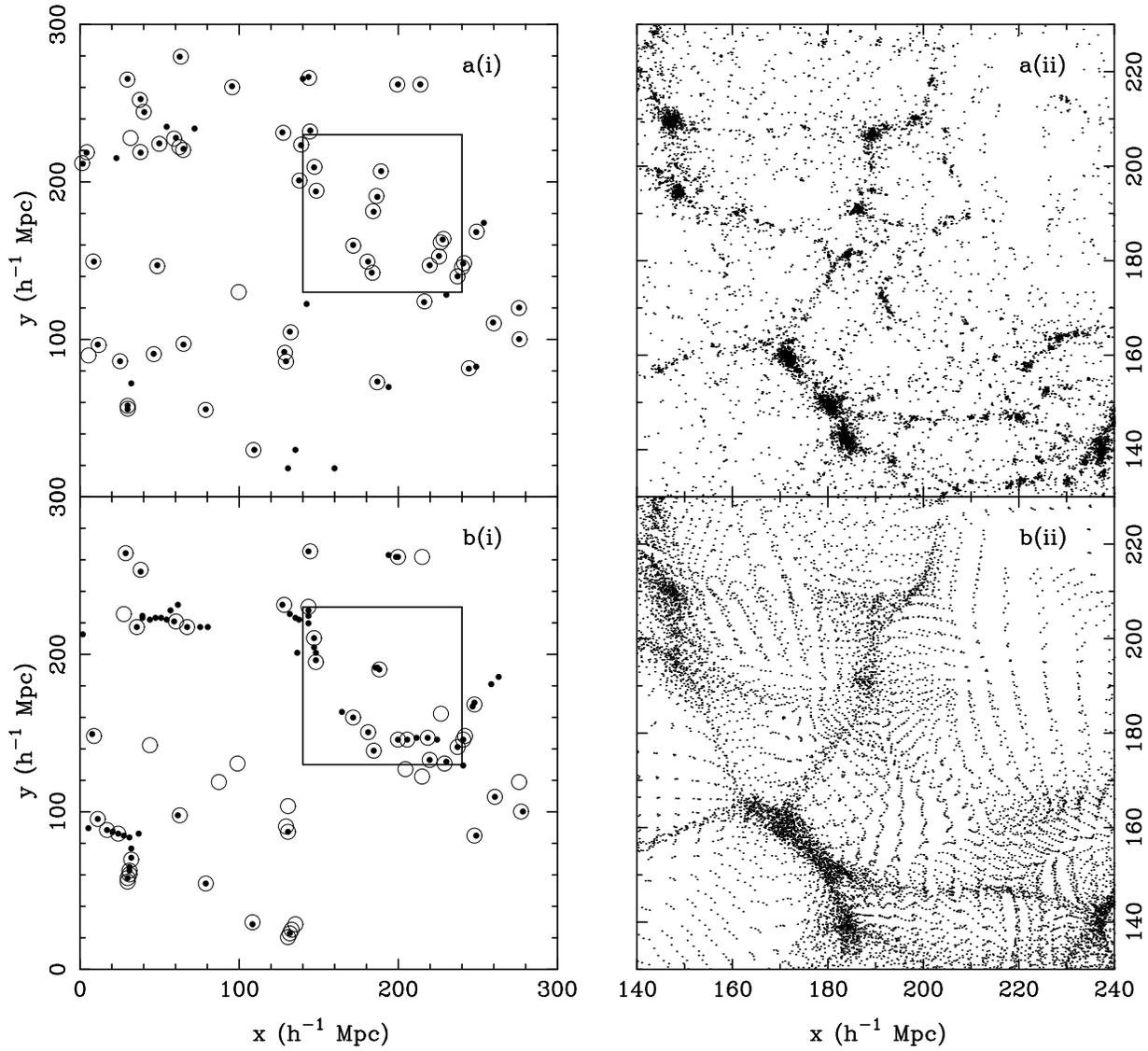

**Figure 1.** A 15 $h^{-1}$ Mpc thickness slice through simulations of $\Gamma = 0.2$ CDM universe with $\sigma_8 = 1.3$ run using (a) a P$^3$M N-body code and (b) the truncated Zel'dovich approximation. The solid dots in panels a(i) and b(i) show the postions of clusters identified in these simulations as high peaks after smoothing the density with a Gaussian filter of radius 1.0 $h^{-1}$ Mpc. The open circles show clusters identified as peaks with Gaussian smoothing of with filter radius 2.0 $h^{-1}$ Mpc in panel b(i) and percolation groups in panel a(i). In panels a(ii) and b(ii) we zoom in on the 100 $h^{-1}$ Mpc square region enclosed by the solid lines in a(i) and b(i), showing the distribution of particles in the N-body and Zel'dovich simulations respectively.

We apply this cluster-finding scheme to the TZA simulation. Because the nature of objects to be identified with clusters is not obvious from the particle plot, we try several different filter sizes $r_f = 1.0, 2.0$ and $3.0$ $h^{-1}$ Mpc. We then look for peaks in the smoothed field. The results for $r_f = 1.0$ $h^{-1}$ Mpc and $2.0$ $h^{-1}$ Mpc are shown in Fig. A1b(i), where we plot those of the 1000 highest peaks which fall in a slice of thickness 15 $h^{-1}$ Mpc. Again it is instructive to compare with the N-body results. We can see that neither value of $r_f$ produces a good correspondence between TZA clusters and the N-body clusters. It also seems that the obvious caustics which the TZA produces are places where maxima are preferentially selected, particularly when the smoothing radius is small. A larger filter scale $r_f$ than with the N-body results should be necessary in order to avoid any particle discreteness problems, as the peak regions do not have such a high particle density. We have run TZA simulations with $200^3$ particles in the same sized box in order to check on the effects of varying mass and spatial resolution. Our recovery of similar results to the tests with $100^3$ particles suggests that the mismatch between TZA and N-body clusters is caused mostly by the TZA dynamical scheme itself being inadequate.

### A3 Clustering statistics and Zel'dovich clusters.

It is possible though that the two sets of objects can give statistically similar results, on the large ($> 10$ $h^{-1}$ Mpc) scales used in comparison with observations, and on which the TZA should be most useful because the density field is more linear. We have calculated the variance $\overline{\xi}_2(r)$ for the TZA and N-body clusters, the results being shown in Fig A2. At this point it is worth mentioning that the method of selecting clusters from a grid (which is necessary in the TZA case) affects the correlations on small scales. The anti-correlation introduced by the fact that peaks must be a minimum of 2 grid cells apart will make $\overline{\xi}_2(r)$ artificially low on small scales. The Gaussian smoothing radius $r_f$ used on the evolved field before peak selection will also depress $\overline{\xi}_2(r)$. This effect is visible in the plot for the N-body results: using larger $r_f$ results in deviations from the percolation curves out to larger r. The TZA clusters should also be affected by both problems, but it difficult to decide from the plot because we do not know what properties a TZA sample with no gridding or smoothing would have. We will therefore concentrate on scales $> 10$ $h^{-1}$ Mpc in the rest of our discussion. The N-body values of $\overline{\xi}_2(r)$ calculated after using percolation and peaks to find clusters are almost identical. The value of $r_f$ which gives the worst agreement, 2.0 $h^{-1}$ Mpc, is probably too large for the compact objects present in the N-body output and in any case using a grid and filter is not our method of choice for selecting N-body clusters. For the TZA clusters, however, the correlation strengths depend strongly on the smoothing radius used to select clusters. It would appear that we are selecting a different population of objects as the smoothing radius is changed. With a small radius there is marked excess clustering, as those peaks in the initial density field which would collapse to form large single clusters are instead broken up into many smaller objects. Using $r_f = 2.0$ gives the closest results to the N-body clusters, although $\overline{\xi}_2(r)$ is still marginally too low. It is important to realise that when comparing a range of Zel'dovich models realistically with observations, we should have an a priori reason for choosing the smoothing scale. The best scale may depend on the shape of P(k) or the amplitude of fluctuations. We have seen here that varying $r_f$ between 1.0 and 3.0 $h^{-1}$ Mpc gives enormously different results for $\overline{\xi}_2(r)$. Also, it apparent from the plots of cluster positions that even with the best radius we are not selecting the same objects as in the N-body output. This may imply that other properties of the cluster density field are not reliably predicted by this method.

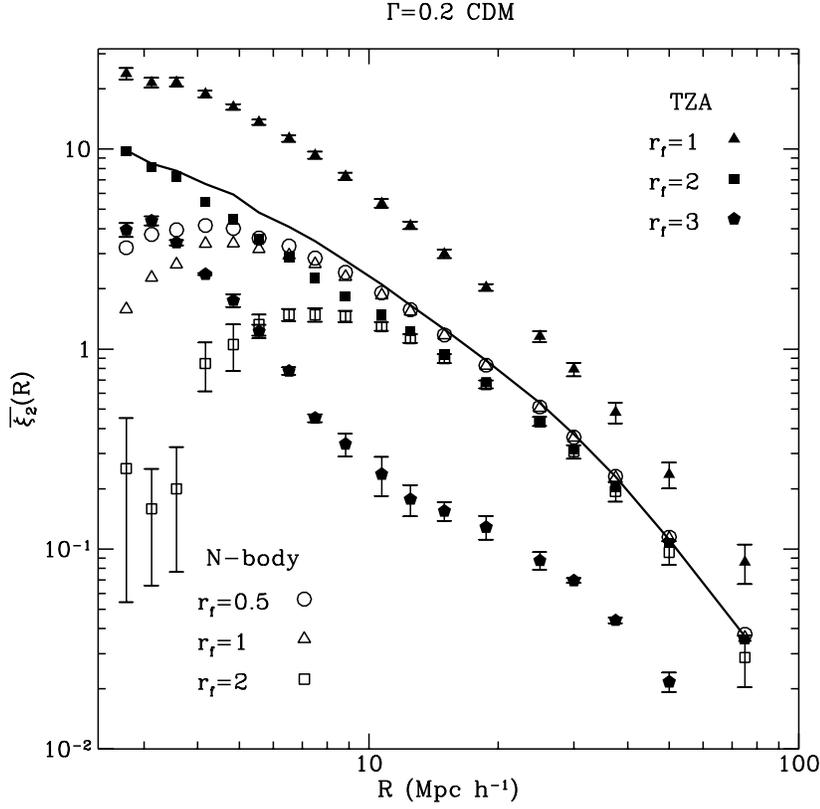

**Figure A1.** The variance $\overline{\xi}_2(r)$ of clusters identified from simulations of a $\Gamma = 0.2$ CDM universe with $\sigma_8 = 1.3$ run using a P³M N-body code and the truncated Zel'dovich approximation. The error bars represent the rms error on the mean calculated from the scatter between 4 realisations. In the N-body results we show $\overline{\xi}_2(r)$ for clusters selected from percolation groups (solid line) and from peaks in the density field (open symbols) smoothed with a Gaussian filter of radius 0.5 , 1.0 and 2.0 $h^{-1}$ Mpc. The filled symbols correspond to the different results obtained after using Gaussian filters of radius 1.0, 2.0, and 3.0 $h^{-1}$ Mpc to smooth the TZA evolved density field. Clusters were identified as the 1000 highest peaks in this field, giving a mean intercluster separation of 30 $h^{-1}$ Mpc.

**A4 Discussion.**

As there is no clear visual correspondence between the N-body and TZA clusters, the exact nature of the objects we are selecting in the TZA density field is unclear. When we use a small $r_f$ or the unsmoothed field, then it is possible that we are affected by particle discreteness, although our tests with higher resolution TZA simulations show that this is not a major problem here. This suggests that some of the objects found in the TZA output are artifacts of the limited dynamics, as are for example the shells of matter that have undergone obvious orbit crossing. By smoothing the initial power spectrum, as we are forced to do, and because the TZA does not work well on small scales, we are losing information which we need to locate where virialised objects will form. This is related to the problem that the efficacy of the TZA will depend on the overall amplitude of fluctuations. For higher normalisations, we will need a larger truncation scale on the power spectrum to suppress the increased orbit crossing. The objects that form will correspond even less to those in the fully non-linear calculation. This implies that the TZA could perform better in comparison with N-body simulations of the other models with different power spectra and lower $\sigma_8$, which we have not tested here. On the other hand, as the galaxy clusters which we are using to trace the density field are by definition compact objects, and probably virialised, our theoretical predictions should be able to predict the location of these objects. The Zel'dovich Approximation, whilst reproducing some other aspects of the large scale evolution of the density field, is unable to do this.

In summary, we find that we can reliably select clumps of matter which we identify as galaxy clusters from N-body simulations, with different selection methods producing almost identical results. This gives us confidence that our theoretical calculations of $\bar{\xi}_J$ will be robust. We have also tested a method for generating simulated cluster catalalogues using the Zel'dovich approximation, and compared the clusters selected using this method with those in our own fully nonlinear calculation. This leads us to conclude that the use of clusters from the Zel'dovich approximation is not really suitable when engaging in a quantitative comparison of theories with observations of cluster clustering.